\documentclass[
    ,final            
  ]
  {aipproc}

\layoutstyle{8x11double}

\usepackage{amsmath,amssymb}
\usepackage{latexsym,graphicx,color,verbatim}
\def\be{\begin{eqnarray}}
\def\ben{\begin{eqnarray*}}
\def\ee{\end{eqnarray}}
\def\een{\end{eqnarray*}}

\def\p{\partial}

\def\=:{=\hspace{-.7em}\raisebox{1.1ex}{.}\hspace{.1em}\raisebox{-0.2ex}{.} }

\newcommand{\NF}{N_{\rm F}}
\newcommand{\NC}{N_{\rm C}}

\newcommand {\beq}{\begin{eqnarray}}
\newcommand {\eeq}{\end{eqnarray}}

\begin{document}

\title{Dynamics of Strings between Domain Walls}
\classification{11.27.+d, 11.25.-w, 11.30.Pb, 12.10.-g}
\keywords      {SUSY, Solitons, BPS}

\author{Minoru Eto}{
  address={ INFN, Sezione di Pisa,
Largo Pontecorvo, 3, Ed. C, 56127 Pisa, Italy}
}

\author{Toshiaki Fujimori}{
  address={Department of Physics, Tokyo Institute of
Technology, Tokyo 152-8551, Japan}
}

\author{Takayuki Nagashima}{
  address={Department of Physics, Tokyo Institute of
Technology, Tokyo 152-8551, Japan} 
}

\author{Muneto Nitta}{
  address={Department of Physics, Keio University, Hiyoshi, Yokohama,
Kanagawa 223-8521, Japan} 
}

\author{\\Keisuke Ohashi}{
  address={Department of Applied Mathematics and Theoretical Physics,
University of Cambridge, CB3 0WA, UK} 
}

\author{Norisuke Sakai}{
  address={Department of Mathematics, Tokyo Woman's Christian University, 
Tokyo 167-8585, Japan } 
}

\begin{abstract}
Configurations of 
vortex-strings stretched between or ending on domain walls 
were previously found to be 1/4 BPS states. 
Among zero modes of string positions, 
the center of mass of strings in each region 
between two adjacent domain walls is shown to be non-normalizable 
whereas the rests are normalizable. 
We study dynamics of vortex-strings stretched 
between separated domain walls by using two methods, 
the moduli space (geodesic) approximation of full {1/4} BPS states 
and the charged particle approximation for 
string endpoints in the wall effective action. 
In the first method 
we obtain the effective Lagrangian explicitly and 
find the 90 degree scattering for head-on collision.
In the second method 
the domain wall effective action is assumed to be 
$U(1)^N$ gauge theory, and 
we find a good agreement between two methods 
for 
well separated strings.
This paper is based on our work \cite{1}.
\end{abstract}

\maketitle


\section{Composite solitons of walls and vortices}
\begin{figure}
\includegraphics[width=120mm]{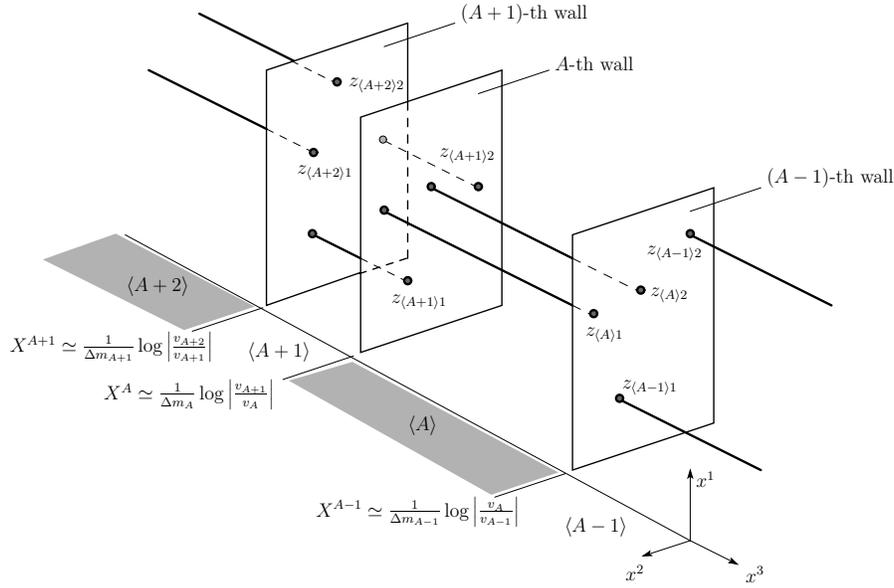}
\caption{Composite solitons of walls and vortices}
\label{dbs}
\end{figure}

Our model is $d=3+1$ $U(1)$ gauge theory with $\NF (\ge 2)$ fundamental Higgs fields
and an adjoint scalar field.
\begin{align}
\mathcal{L} &= - \frac{1}{2g^2} F_{\mu \nu}F^{\mu \nu} 
+ \frac{1}{g^2} \p_\mu \Sigma \p^\mu \Sigma 
+ \mathcal D_\mu H (\mathcal D^\mu H)^\dag \notag \\
&- \frac{g^2}{4} \left(HH^\dag -c \right)^2 
+ (HM-\Sigma H)(HM-\Sigma H)^\dag .
\end{align}
The model can embedded into a ${\cal N}=2$ supersymmetric gauge theory
with non-zero FI parameters. Vacua of this model are characterized by
one flavor index $\langle A \rangle$
\beq
H^A=\sqrt{c}, \quad H^B=0\,\, (B \neq A), \quad \Sigma=m_A.
\eeq
There are $\NF$ discrete vacua labeled by $\langle A \rangle$,
and there exist 1/2 BPS domain walls
dividing them \cite{walls}.
Furthermore, vortices can live in each vacuum breaking further 1/2 supersymmetries.
The 1/4 BPS equations can be obtained as\footnote{
These equations are obtained in $U(\NC)$ gauge theory.
If we consider $U(\NC \ge 2)$ gauge symmetry, there appear monopoles with flux tubes.
However, we concentrate on Abelian gauge theory and
consider composite solitons of walls and vortices in this talk.}\cite{walls and vortices}
\beq
 {\cal D}_1 \Sigma - F_{23} =0,
\quad 
  {\cal D}_2 \Sigma - F_{31} =0,
\\
 {\cal D}_3 \Sigma - F_{12} 
- \frac{g^2}{2}\left(c{\bf 1}_{N_{\rm C}} - HH^{\dagger}\right)  =0 ,
\\
 {\cal D}_1 H + i  {\cal D}_2 H =0, 
\quad 
 {\cal D}_3 H +  \Sigma H-H M=0.
\eeq
Configurations of vortices and domain walls are illustrated in Fig.\ref{dbs}.

\section{Dynamics of vortices between domain walls}
Dynamics of solitons can be investigated by moduli space approximation \cite{Manton}.
We give the weak time dependence to normalizable moduli parameters,
and look at the time evolution
of them. Geodesic motions on the moduli space correspond to
actual motion of solitons.
Here let us focus on $\NF=3$ with masses $M=(m,0,-m)$ and consider
the configuration with two domain walls
and two vortices in the middle vacuum.
Such exact solution can be obtained
 in strong coupling limit $g\to \infty$.
\begin{figure}[htb]
\begin{tabular}{ccc}
\includegraphics[height=4cm]{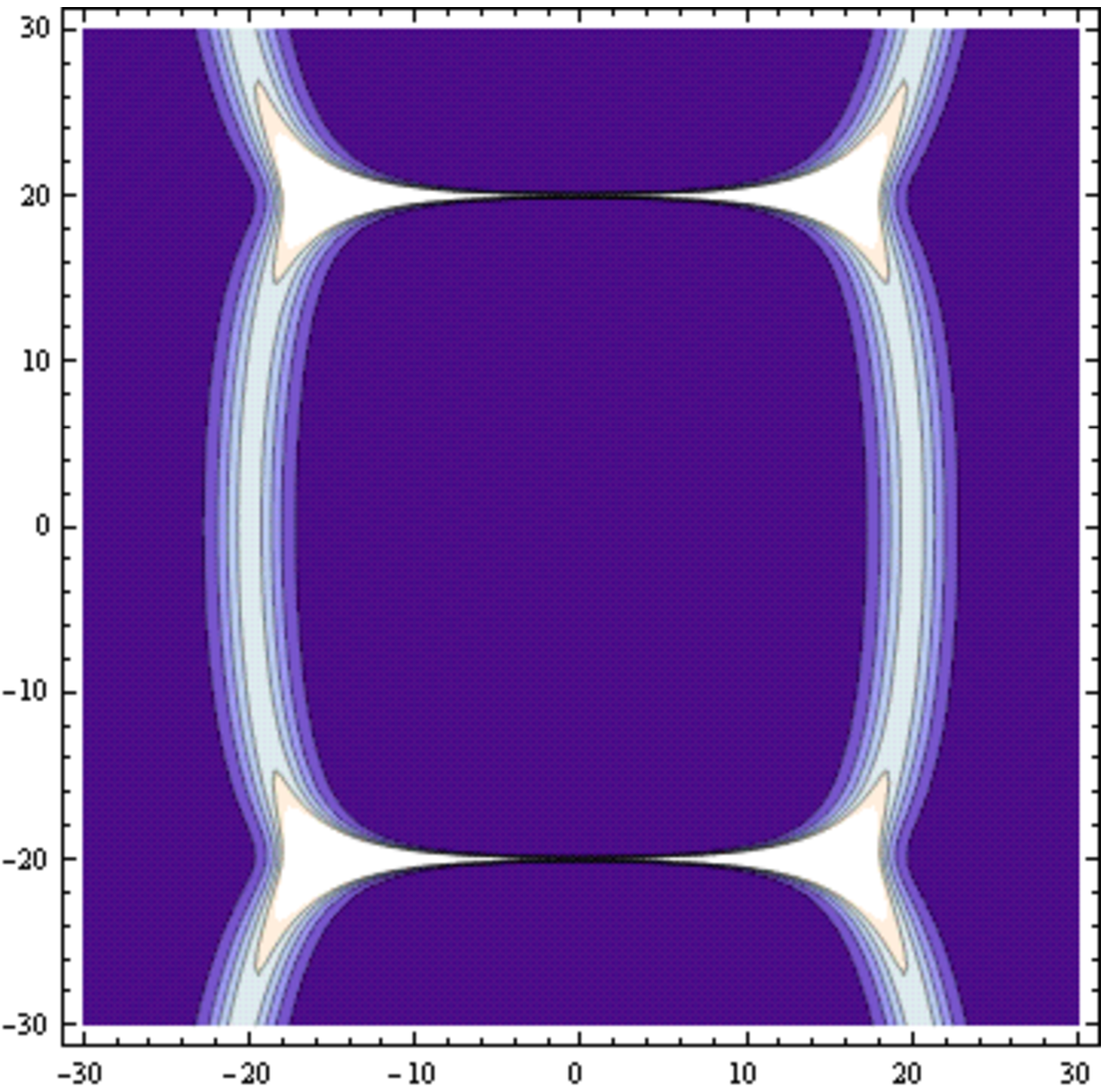} &
\includegraphics[height=4cm]{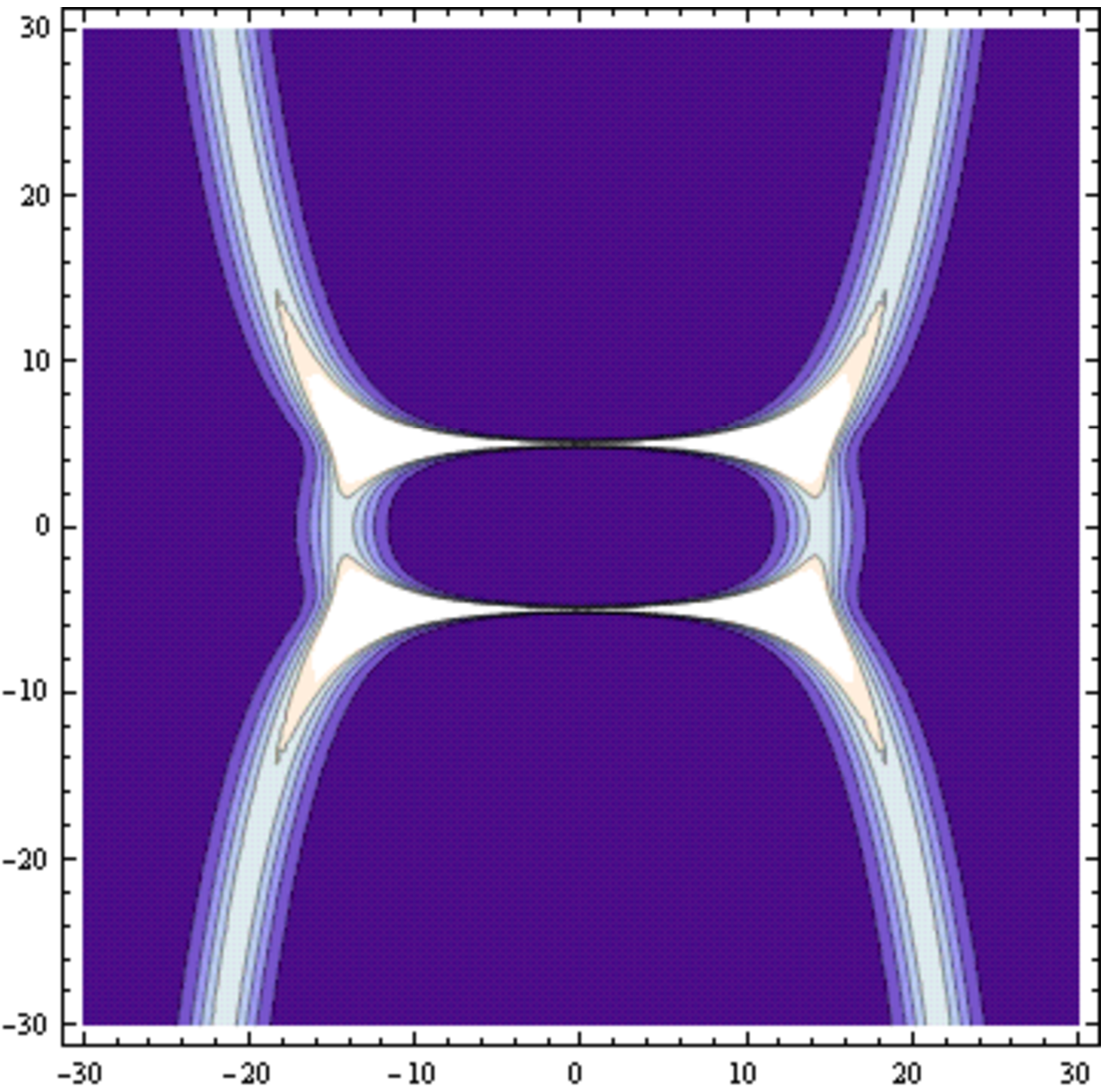} &
\includegraphics[height=4cm]{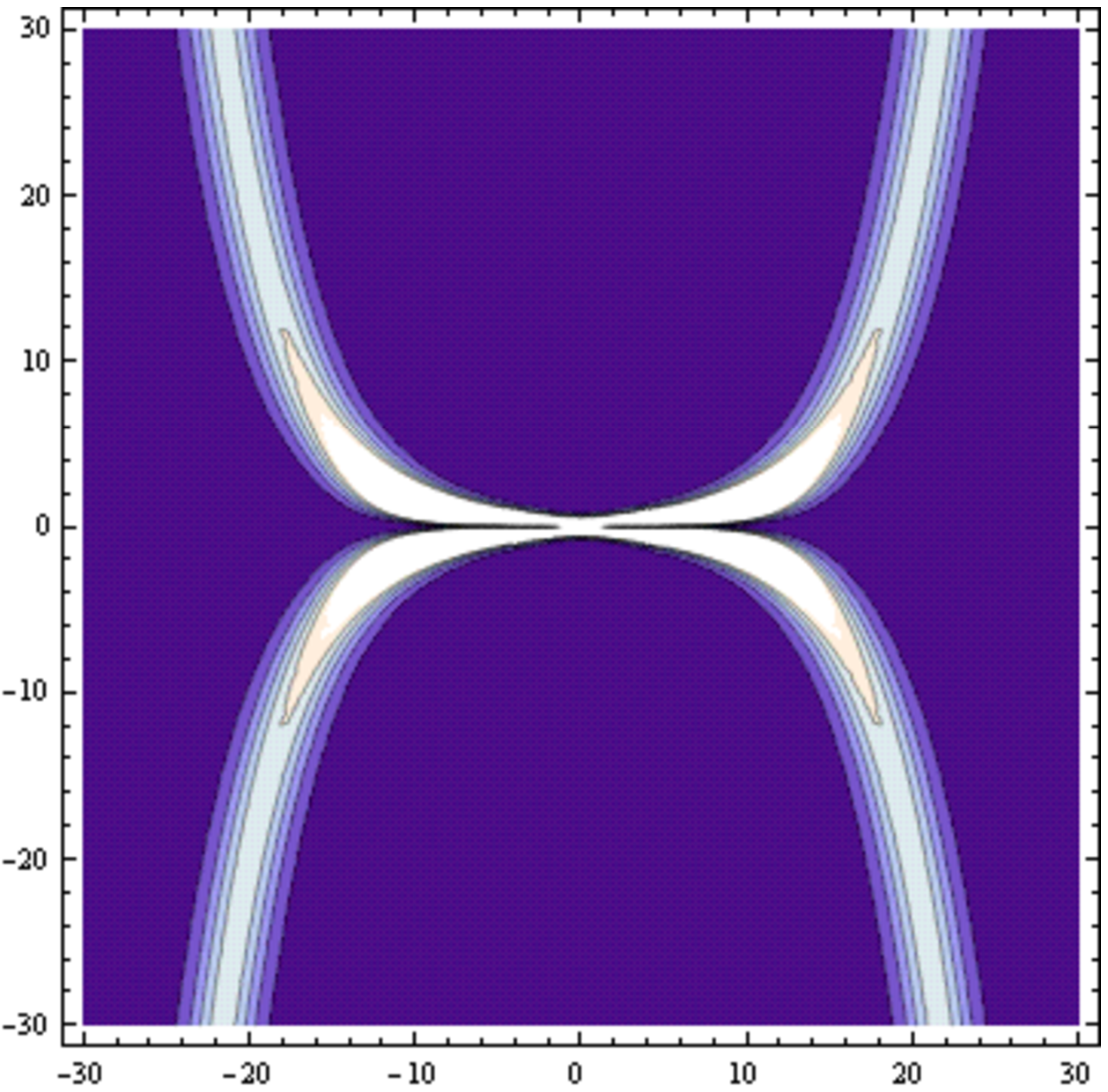}\\
$z_0=20$ & $z_0=5$ 
& $z_0=0$
\end{tabular}
\caption{The energy densities in a plane containing vortices.
Vertical lines are walls and horizontal lines are vortices.}
\label{020}
\end{figure}
Normalizable moduli parameter in the solution is only relative distance between
two vortices $z_0$. Distance between domain walls is controlled by
non-normalizable moduli parameter $v$.
The energy densities in a plane containing vortices with
various $z_0$ and $v=e^4$ are shown in Fig.\ref{020}.

Let us now give the weak time dependence to $z_0$, 
and investigate the dynamics of the middle vortices. 
We can obtain the K\"ahler metric as 
an integral over the complete elliptic integral 
of the second kind $E(k)$ 
\begin{align}
K_{z_0 \bar z_0}&=2 \pi c \int dx_3 \, k E(k),\notag \\
& \quad {\rm with} \quad 
k=\left(\frac{|v z_0^2|^2}{2 \cosh mx_3 + |v z_0^2|^2} \right)^{1/2}. 
\end{align}
If we expand the K\"ahler metric around $|v z_0^2|^2 = 0$, we obtain
\begin{align}
dx^2 &= 2 K_{z_0 \bar z_0} d z_0 d \bar z_0 \notag \\
&= \frac{\pi^{\frac{3}{2}}|v|c}{4m} 
\Bigl( \Gamma(1/4)^2 - \frac{3}{2} \Gamma(3/4)^2 |v Z|^2  \notag \\
& \qquad \qquad \qquad \qquad + \mathcal O(|v Z|^4) \Bigr) d Z d \bar Z.
\end{align}
Since the coordinate $Z \equiv z_0^2$ is a good coordinate 
even at the origin, 
it shows that the moduli space is non-singular at the origin and 
the vortices scatter with right-angle in head-on collisions. 
If we take the opposite limit $|v z_0^2|^2 \to \infty$, 
the metric can be calculated as 
\beq
K_{z_0 \bar z_0} \approx \frac{8 \pi c}{m} \log |4v z_0^2|. 
\label{eq:asym_metric_020}
\eeq
Since the domain walls are logarithmically bending in the present case, 
the definition of the distance between domain walls 
is not clear. However, at the center of mass of two vortices, 
the distance between domain walls is given by 
\beq
l_{\langle 2 \rangle} = \frac{4}{m} \log |v z_0^2|.
\label{eq:020-length}
\eeq
It can be considered as 
the typical lengths of the vortices (see Fig.\ref{020}). 
Therefore, the above asymptotic metric (\ref{eq:asym_metric_020}) 
can be understood as the kinetic energy of two vortices. 

\section{Vortices as charged particles}
So far, we have calculated 
the metric on the 1/4 BPS moduli space and 
investigated the dynamics of vortices suspended 
between the domain walls, 
using the moduli space approximation. 
Now let us obtain the vortex dynamics from effective theory on domain walls. 
Effective theory on $N$ domain walls is given by
$N$ scalar fields $\Phi_A$ and $N$ compact scalar fields $\sigma_A$,
which correspond to positions and phases of domain walls, respectively.
If we take the dual of compact scalar fields in $d=2+1$, we obtain $U(1)^N$ gauge
theory as the effective theory on domain walls
\begin{eqnarray}
 {\cal L}_{\rm w}=
\sum_{A=1}^N\left[\frac{1}{2e^2_A}\partial_\mu \Phi^A \partial^\mu \Phi^A
-\frac{1}{4e^2_A}F_{\mu \nu}^AF^{A\,\mu \nu}\right].
\label{eq:multi_dual}
\end{eqnarray}
Here we assume $N$ domain walls are all well-separated.
Vortices can be viewed as charged particles in the effective theory, which are
sources of scalar fields and electric fields
on the neighboring domain walls.
Let us consider the $i$-th vortex living in vacuum $\langle B \rangle$ 
positioned at $z_{\left<B\right>i}$ with a velocity 
$\dot{z}_{\left<B\right>i}=u_{\left<B\right>i}$. It yields 
the scalar field and the electric field on the worldvolume of 
neighboring domain walls
\begin{align}
(\Phi^A)_{(B,i)} &= 
\phantom{-} (\delta_{A+1,B} - \delta_{AB}) \frac{e^2_A}{2\pi} 
G(u_{\left<B\right>i};z-z_{\left<B\right>i}), \label{eq:Phi} \\
(A_\mu^A)_{(B,i)} &=
- (\delta_{A+1,B} - \delta_{AB}) \frac{e^2_A}{2\pi} 
\frac{v_{\mu\left<B\right>i}}{\sqrt{1-|u_{\left<B\right>i}|^2}}\notag \\
& \qquad \qquad \times G(u_{\left<B\right>i};z-z_{\left<B\right>i}). \label{eq:A}
\end{align}
where $v^\mu_{\left<B\right>i}=(1,{\rm Re}[u_{\left<B\right>i}], {\rm Im}[u_{\left<B\right>i}])$ 
and $G$ is the Green's function given by 
\begin{align}
 G(u_{\left<B\right>i};z-z_{\left<B\right>i})=\log|L_{u_{\left<B\right>i}}(z-z_{\left<B\right>i})|
-\log L+f(u_{\left<B\right>i}).
\end{align}
We can regard 
the dynamics of the vortex living in vacuum 
$\langle A \rangle$ as an 
electric charge moving in the background potential 
produced by the other vortices.
The Lagrangian for the $i$-th particle in vacuum $\langle A \rangle$ 
is then given by
\begin{align}
L_i^{\langle A \rangle} &= \sum_B (\delta_{A,B} - \delta_{A-1,B}) 
\left( - \widetilde \Phi^B \sqrt{1-|u_{\left<A\right>i}|^2} - \widetilde A_\mu^B v^\mu_{\left<A\right>i} \right) \nonumber \\
&\approx \sum_B (\delta_{A,B} - \delta_{A-1,B}) 
\left( - \widetilde \Phi^B + \frac{\widetilde \Phi^B}{2} 
|u_{\left<A\right>i}|^2 - \widetilde A_\mu^B v^\mu_{\left<A\right>i} \right),
\label{eq:charge_lag}
\end{align}
where $\widetilde \Phi^B,\, \widetilde A_0^B,\, \widetilde {\mathbf A}^B$ are the values of the fields produced by the other particles at the location of the particle $z=z_{\left<A\right>i}$
\beq
\widetilde \Phi^B &\equiv& \langle \Phi^B \rangle + \sum_{(C,j)} (\Phi^B)_{(C,j)} \big|_{z=z_{\left<A\right>i}}, \\
\widetilde A_\mu^B &=& \sum_{(C,j)} (A_\mu^B)_{(C,j)} \big|_{z=z_{\left<A\right>i}}.
\eeq
Similar method is well-known for monopoles in $3+1$ dimensions \cite{monopoles}.

The above procedure yields
the effective Lagrangian for the relative motion 
of two vortices which we have investigated in the last section as 
\beq
L_{\rm eff} = \left( \frac{8 \pi c}{m} \log
|v z_0^2| + \frac{16 \pi c}{m} \log 2 \right) |\dot{z}_0|^2.
\eeq
This coincides with the asymptotic result 
in 
Eq.(\ref{eq:asym_metric_020}).

Here we have not shown other examples. However,
we can show that this method correctly reproduces 
the asymptotic metric on the moduli space when 
the domain walls are well-separated in $x_3$-direction, 
and the 
vortices
are 
well-separated 
from other vortices in $z$-plane.





\bibliographystyle{aipproc}   

\bibliography{sample}

\IfFileExists{\jobname.bbl}{}
 {\typeout{}
  \typeout{******************************************}
  \typeout{** Please run "bibtex \jobname" to optain}
  \typeout{** the bibliography and then re-run LaTeX}
  \typeout{** twice to fix the references!}
  \typeout{******************************************}
  \typeout{}
 }

\end{document}